\documentclass[prb,twocolumn,floats,superscriptaddress,citeautoscript]{revtex4}

\usepackage{graphicx}
\usepackage{amsmath}
\usepackage{amssymb}
\usepackage{times}
\usepackage{colordvi}
\usepackage{textcomp}

\graphicspath{{.}{./EPS/}}

\begin{document}

\title{Universal characteristics of resonant--tunneling field emission from
nanostructured surfaces}

\author{S. Johnson}
\affiliation{Institute of Geological and Nuclear Sciences, PO Box 31-312,
30 Gracefield Road, Lower Hutt, New Zealand}

\author{U. Z\"ulicke}
\affiliation{Institute of Fundamental Sciences, Massey University, Private Bag
11~222, Palmerston North, New Zealand}
\affiliation{MacDiarmid Institute for Advanced Materials and Nanotechnology,
PO Box 600, Wellington, New Zealand}

\author{A. Markwitz}
\affiliation{Institute of Geological and Nuclear Sciences, PO Box 31-312,
30 Gracefield Road, Lower Hutt, New Zealand}
\affiliation{MacDiarmid Institute for Advanced Materials and Nanotechnology,
PO Box 600, Wellington, New Zealand}

\date{\today}

\begin{abstract}

We have performed theoretical and experimental studies of field
emission from nanostructured semiconductor cathodes. Resonant tunneling
through electric--field--induced interface bound states is found to strongly
affect the field--emission characteristics. Our analytical theory predicts
power--law and Lorentzian--shaped current--voltage curves for resonant--tunneling
field emission from three--dimensional substrates and two--dimensional
accumulation layers, respectively. These predicted line shapes are observed in
field emission characteristics from self--assembled silicon nanostructures. A
simple model describes formation of an accumulation layer and of the resonant
level in these systems.

\end{abstract}


\maketitle
\section {Introduction}
Field emission describes the ejection of electrons from a conducting substrate
into vacuum via electron tunneling through the surface potential
barrier~\cite{fowl:procRS:28,nord:procRS:28}.  As tunneling is a sensitive
probe of the electronic spectrum, field emission has often been applied as a
useful spectroscopic tool for bulk--material characterization~\cite{plum:rmp:73}.
Experiments using {\em nanostructured\/} emitters~\cite{lin:prl:91,garcia:prl:92}
reveal deviations from conventional
theory~\cite{fowl:procRS:28,nord:procRS:28}, signifying that, in these systems,
tunneling occurs through discrete energy levels formed in the nanotips due to
spatial confinement. In addition to revealing intriguing properties of electron
transport at the nanoscale, the intense and highly coherent electron beams
extracted from such nanotips have found useful applications in projection
electron holography~\cite{fink:prl:90}. More recently, a number of field--emission
studies of 2D (quantum film), 1D (quantum wire)~\cite{lito:ass:97,lito:jap:04},
and 0D (quantum dot)~\cite{naum:pss:03}\ {\em semiconductor\/} systems have
also demonstrated clear deviations from Fowler--Nordheim behavior. In
particular, discrete current peaks, an anomalously low threshold field for electron
emission and regions of negative differential conductance have all been
observed in the current--voltage ({\it I--V}) field--emission characteristics. All of
these features are associated with the formation of quantized bound states in the
low--dimensional semiconductor cathodes. These experimental results have
stimulated a number of theoretical studies of electron field emission from and
through quantum--confined electronic
states~\cite{lito:jvac:96,huang:jap:95,vat:jap:97,gonch:mat:03,wang:prb:05}.

\begin{figure}
\includegraphics[width=2.9in]{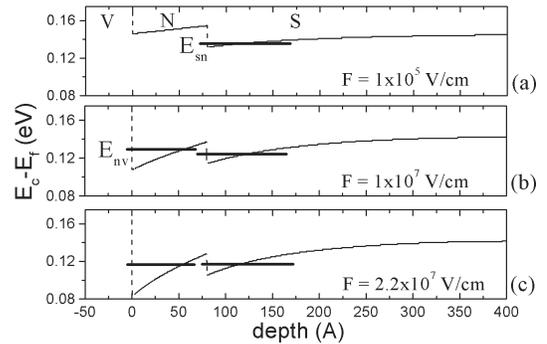}
\caption{Energy band diagram of a realistic model for a
vacuum-nanostructure--substrate system based on a silicon nanostructure 8 nm
high and 10 nm in diameter. The external electric field $F$ is increasing
in successive panels (a)--(c). V, N, and S correspond to the vacuum,
nanostructure and substrate regions, while $E_{\text{sn}}$ and
$E_{\text{nv}}$ correspond to the lowest energy bound states located at the
substrate--nanostructure and nanostructure--vacuum interfaces respectively.
\label{Fig1}}
\end{figure}
In this work, we consider a generic system consisting of a bulk semiconductor
substrate (S) with a nanostructured surface where nanometer--sized tips
(`whiskers') have formed. In addition to the energy barrier between electronic
states in the nanostructure (N) and vacuum (V), quantum confinement gives rise
to a further barrier at the substrate--nanostructure (S--N) interface. Electric--field
penetration into the semiconductor material leads to band
bending~\cite{tsong:surf:79,wang:apl:02} and the formation of 2D bound
states at both interfaces. The evolution of bound--state formation is
illustrated in Fig.~\ref{Fig1} for a set of model parameters. We focus on
the situation shown where, initially, a 2D bound state is formed at the S--N
interface. See Fig.~\ref{Fig1}(a). As its energy lies below the substrate 
conduction--band edge, no resonant tunneling can occur via this state. As the
electric field is increased further, quantization of electron energy also occurs
within the potential well formed at the N--V interface, as shown in
Fig.~\ref{Fig1}(b). The corresponding bound states can mediate
resonant tunnelling both from the 3D substrate electron system and also from
the 2D bound states formed in the S--N quantum well. Here we present results
from our analytical theory which describes resonant--tunneling field emission
from such 3D and 2D systems. These studies provide new insight into universal
features exhibited in field emission from nanostructured surfaces and which are
confirmed by our measurements of field--emission currents from self--assembled 
silicon nanostructures. In particular, resonant tunneling from a 3D
substrate via a 2D interface bound state is found to give rise to a power law
$I\propto V^{5/3}$ in the ascending part of the resonance peak in the {\em
I--V\/} curve. The clear contrast to Fowler--Nordheim theory, which predicts $I
\propto V^2$,  allows for unambiguous identification of the 3D--vacuum 
resonant--tunneling mechanism. Alternatively, when resonant tunneling 
through the 2D interface bound state occurs from a 2D accumulation layer,
the peak exhibits a Lorentzian line shape. We start by describing the
theoretical method and results. This is followed by details of our sample
preparation, measurement setup, and the obtained field--emission data.

\section {Theoretical analysis}
The general expression for the field--emission current density from a cathode
as a function of temperature $T$ and voltage $V$ is given 
by~\cite{daviesBook,gonch:mat:03}
\begin{equation}\label{genCurrent}
j_{\text{FE}}(T,V)=\frac{e}{\pi\hbar} \int_0^\infty \!\! dE \,\,\, T(E,V)\,
N(E,T,V) \quad .
\end{equation}
Here $T(E,V)$ denotes the energy--dependent quantum--mechanical
transmission function from electrode to vacuum, and $N(E,T,V)$ is the
supply function of the electron reservoir, i.e., the density of electrons in
transverse plane--wave states having energy $E$ in their motion
perpendicular to the interfaces. The substrate-vacuum voltage drop $V$ is
proportional to the applied electric field $F$, with the constant of proportionality
being dependent on the specific experimental geometry and system under
consideration. We proceed now to discuss the form of the
transmission and supply functions employed in our transport calculations
before stating the results of our analytical theory.

Resonant tunneling through a 2D bound state, e.g., the one formed at the
N--V interface shown in Fig.~\ref{Fig1}, can be described
phenomenologically~\cite{daviesBook} by a Lorentzian energy dependence
of the transmission coefficient,
\begin{equation}
T_{\text{RT}}(E)=T_{\text{pk}}\left[1+4\left(\frac{E-E_{\text{nv}}}
{\Gamma_{\text{nv}}}\right)^2\right]^{-1} \quad ,
\end{equation}
where $E_{\text{nv}}$ and $\Gamma_{\text{nv}}$ are the resonance energy
and life--time broadening, respectively, of the bound state at the N--V
interface. $T_{\text{pk}}$ is the transmission probability at resonance. As the
double--barrier system under consideration is highly asymmetric, with a high
(low) barrier at the N--V (S--N) interface, the peak transmission is dominated
by Fowler--Nordheim tunneling~\cite{FNcaveat} through the N--V barrier.
The relation 
\begin{equation}
\frac{\pi}{2} \, T_{\text{pk}} \, \Gamma_{\text{nv}} \approx E_{\text{nv}} \,
T_{\text{nv}}(E_{\text{nv}})
\end{equation}
can then be derived~\cite{daviesBook},  with the transmission function for
tunneling through the N--V barrier denoted by $T_{\text{nv}}(E)$.

The explicit form of the supply function depends on the dimensionality of the
electron reservoir. For field emission from a 3D substrate, it is given by the
familiar expression~\cite{daviesBook}
\begin{equation}
N_{\text{sub}}(E,T,V)=\frac{m k_{\text{B}} T}{2\pi\hbar^2}\,\ln\left( 1 + 
e^{\frac{\mu_{\text{s}} + e V - E}{k_{\text{B}} T}}\right) \quad ,
\end{equation}
with $k_{\text{B}}$ denoting the Boltzmann constant, $m$ being the effective
mass of electrons in the substrate, and $\mu_{\text{s}}$ the equilibrium
chemical potential of the substrate. The supply function is modified, however,
when a 2D system such as the accumulation layer formed at the S--N
interface acts as the electron reservoir. While the energy of electron motion
transverse to interfaces is quantized in a quantum well, coupling of the bound
state to substrate and nanostructure lead to a finite life time of electrons in
the accumulation layer. Hence it supplies electrons with energies in a narrow
range $\Gamma_{\text{sn}}$ around the bound--state energy $E_{\text{sn}}$
with a Lorentzian distribution~\cite{gonch:mat:03}, yielding
\begin{equation}
N_{\text{acc}}(E,T,V)=N_{\text{sub}}(E,T,V) \left[1+4\left(\frac{E-E_{\text{sn}}}
{\Gamma_{\text{sn}}}\right)^2\right]^{-1} \, .
\end{equation}

Having specified transmission and supply functions, we are in the position to
calculate the field--emission characteristics using Eq.~(\ref{genCurrent}). To
obtain our analytical results, we make the following assumptions:
(i)~presence of a highly degenerate electron gas in the reservoir, which
means that we use the $T\to 0$ limiting expression for the supply functions,
and (ii)~triangular shape of the N--V quantum well, which allows us to
express $E_{\text{nv}}$ in terms of the applied electric field $F$
as~\cite{daviesBook} $E_{\text{nv}}=(\hbar^2 \pi^2/2 m \varepsilon^2)^{1/3}
(e F)^{2/3}$. We find then the following expressions for the {\it I--V\/}
characteristics of resonant--tunneling field emission from a 3D substrate and
from a 2D accumulation layer, respectively:
\begin{subequations}\label{results}
\begin{widetext}
\begin{equation}\label{3Dcase}
j_{\text{RTFE}}^{\text{(3D)}}(V) = \frac{e^2 m}{2\pi^2\hbar^3} \left( \frac
{\pi^2 \hbar^2}{2 m\kappa D^2}\right)^{\frac{1}{3}} \left( e V \right)^{\frac{5}{3}}
\exp\left[ -\frac{V_0}{V}\right] \, \theta \left( \mu_{\text{s}} - E_{\text{nv}} \right)
\theta \left( E_{\text{nv}} - E_{\text{bs}}\right) \quad ,
\end{equation}
\end{widetext}
\begin{equation}\label{2Dcase}
j_{\text{RTFE}}^{\text{(2D)}}(V) = j_{\text{RTFE}}^{\text{(3D)}}(V) \,\, \frac
{\Gamma_{\text{sn}}}{\Gamma_{\text{sn}}+\Gamma_{\text{nv}}} \,\, \frac
{\theta\left(\mu_{\text{s}} - E_{\text{sn}} \right)}{1 + \left[ 2 \frac{E_{\text{sn}}
- E_{\text{nv}}}{\Gamma_{\text{sn}}+\Gamma_{\text{nv}}} \right]^2} \quad .
\end{equation}
\end{subequations}
Here $\kappa$ denotes the dielectric constant of the substrate material, $D$
is the distance over which the cathode potential is dropped, and $V_0$ is a
voltage scale related to the effective work function of the vacuum barrier,
which is modified due to energy quantisation. The Heaviside step functions
$\theta(x)$ in Eqs.~(\ref{results}) enforce proper phase--space conditions for
resonant--tunneling transport to occur, e.g., cutting the current off sharply
when the (in general voltage--dependent) resonant--level energy 
$E_{\text{nv}}$ falls below the conduction--band bottom $E_{\text{bs}}$. The
conspicuous difference in the voltage--dependent pre-exponential factor of
Eq.~(\ref{3Dcase}) compared to the Fowler--Nordheim formula [$(eV)^{5/
3}$ as opposed to $(eV)^2$] arises from the field dependence of the
bound--state energy $E_{\text{nv}} \propto (V/D)^{2/3}$ formed in the
triangular N--V quantum well.

\section {Experimental procedure}
We now proceed to describe details of our experimental study which examines
the field--emission properties of a self--assembled silicon nanostructure formed
on a silicon substrate~\cite{steve:apl:04}. Detailed analysis of the {\it I--V\/}
field--emission characteristics indicates the presence of two classes of current
peaks which occur at different regions of electric field. Based on our
understanding gained from theoretical analysis described above, we can relate
these two classes of current peaks to resonant tunnelling from the 3D
substrate and from 2D bound states located at the N--S interface,
respectively. In both cases, tunnelling is mediated by a second set of 2D
bound states located at the N--V interface. 

\begin{figure}
\includegraphics[width=3in]{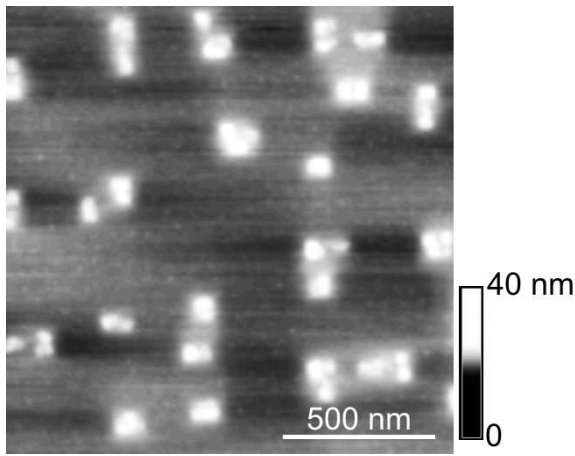}
\caption{1.5 $\mu$m$\times 1.5 \mu$m AFM surface plot of the Si
nanostructures formed by electron beam annealing at 1000 \textcelsius \ for
15 s. The rate at which the sample was heated and cooled was
5 \textcelsius/s.
\label{Fig2}}
\end{figure}
Substrates used in the present study were n-type Si (100) wafers (P-doped,
conductance 1--10 $\Omega$cm , corresponding to a dopant density of
$10^{15}$--$10^{16}$ cm$^{-3}$). Nanostructure growth results from annealing 
of the untreated substrates to 1000 \textcelsius\ for 15 s using a 20 keV
raster--scanned electron beam. Details of the fabrication procedure and
mechanism for nanostructure growth have been reported
elsewhere~\cite{steve:jap:04}. A characteristic AFM image of the nanostructured
Si substrate surface is shown in Fig.~\ref{Fig2}. The self-assembled
nanostructures are square--based and distributed randomly over the surface
with a density of 11 $\mu$m$^{-2}$. High--resolution AFM and TEM
studies~\cite{steve:jap:05} indicate an average nanostructure height of 8 nm
and a base length distribution in the range 8--60 nm. The relevance of these
structures to the study of 3D and 2D resonant--tunnelling field emission 
phenomena of the type discussed theoretically above, becomes apparent 
when one considers an energy band diagram of the S--N--V system.
In the first case, the S--N interface was modelled by an offset in the conduction
band equal to the confinement energy of the lowest--lying transverse bound
state in the nanostructure. Self-consistent solutions of the Poisson and
Schr{\"o}dinger equations were then used to calculate the
conduction--band--bottom profile and to find regions of electron energy
quantization~\cite{snid:jap:90}. Given the size distribution of the nanostructures,
and assuming a rectangular quantum well, the offset in the conduction band was
calculated to be in the range 0.6--31 meV with bound states located at the S--N
interface forming only in those structures smaller than 12 nm.
Fig.~\ref{Fig1} shows the results of our simulations for a 10 nm diameter
silicon nanostructure.

\section {Results and Discussion}
\begin{figure}[t]
\includegraphics[width=3in]{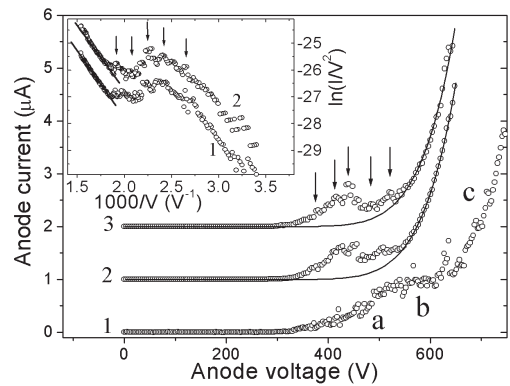}
\caption{{\it I--V\/} field emission characteristics for Si nanostructures formed
on n-type Si (100) substrate. The three curves correspond to successive
conditioning cycles, where 1 corresponds to the first and 3 the final conditioning
cycle . The solid lines that overlay curves 2 and 3 correspond to a fit to
Fowler--Nordheim theory. For clarity, each curve has been offset by 1 $\mu$A .
The inset shows Fowler-Nordheim plots of curves 1 and 2. The arrows
indicate the positions of the five most prominent peaks, while the solid lines
indicate the best fit for the experimental data in the Fowler--Nordheim
emission region. Curve 2 has been offset by 1 along the y-axis.
\label{Fig3}}
\end{figure}
Characterisation of the field--emission properties of these self-assembled 
nanostructures was performed
at room temperature
in a parallel--plate diode configuration. Electrical
isolation and cathode--anode separation was achieved using a 100 $\mu$m
thick PTFE film, and measurements were performed under vacuum of $1\times
10^{-7} $mbar. It was found necessary to run several emission cycles at high
emission currents (5 $\mu$A ) in order for the {\it I--V\/} characteristics to
become stable and reproducible. The measured anode voltage can be related
to the voltage drop $V$, defined in the equations above, via a lever-arm
relationship. The {\it I--V\/} curves plotted in Fig.~\ref{Fig3} show 
the change in emission current during the course of a conditioning cycle. The
nanostructured substrate was maintained under vacuum between successive
cycles. Following incomplete conditioning (curve~1) the emission
characteristics are typified by two regions of exponentially rising current
(labelled a and c, respectively), separated by a plateau in emission current
(region~b). With continued conditioning however, distinct current peaks
develop within the current plateau region (see curves~2 and 3 of
Fig.~\ref{Fig3}). The arrows indicate the positions of the five most
prominent peaks. While the peak positions and relative intensities were found 
to be repeatable for each sample, a significant deviation in the number,
position and intensity of peaks was observed between samples. While it is not possible to identify the individual nanostructure involved in electron emission, AFM images of the substrate surface, taken before and after field emission measurements, appear identical. Further, following
exposure to air, it was found necessary to again condition the cathodes in
order for the current peaks to develop, suggesting that conditioning is related
to removal of adsorbates from the cathode surface.

The current peaks are also clearly observed when plotting the {\it I--V\/}
characteristics on a Fowler-Nordheim plot [$\ln(I/V^2)$ versus $1/V$], as
shown in the inset of Fig.~\ref{Fig3} for curves~1 and 2. The
Fowler--Nordheim plot also reveals highly linear behaviour at anode voltages
$> 550$ V, suggesting electron emission through conventional (direct,
non-resonant) Fowler--Nordheim tunnelling. The solid lines in 
Fig.~\ref{Fig3}, which display the Fowler--Nordheim relation fitted to the
experimental data in this high voltage region, demonstrate excellent
agreement between the model and observed emission current. Current
peaks in the {\it I--V\/} characteristics of substrates containing multiple
cathodes could be related to the activation and de-activation of additional
emission sites~\cite{hasko:jvac:97}. However, the leading edge of each peak
marked in the inset of Fig.~\ref{Fig3} is found to be highly non-linear,
suggesting conduction through a mechanism other than conventional
Fowler--Nordheim tunnelling, in contrast to what would be expected from the
activation of additional emission sites. This is in agreement with
measurements performed using a phosphor--coated anode, which suggests
electron emission from a {\em single emission site\/} over the range of anode 
voltage considered.

\begin{figure}
\includegraphics[width=3in]{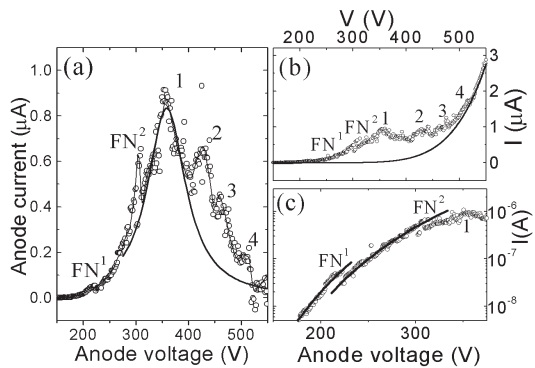}
\caption{(a)~Field emission resonance region following subtraction of the
background Fowler-Nordheim emission current as fitted to the {\it I--V\/}
characteristics displayed in (b). (c)~Semi--logarithmic plot of experimental
data shown in (b). Solid lines indicate the best fits to current peaks FN$^1$
and FN$^2$.
\label{Fig4}}
\end{figure}
In order to study the current peaks in more detail, the background current
fitted to the high--voltage Fowler--Nordheim tunnelling region was subtracted
from the experimental data, as shown in Fig.~\ref{Fig4}(a). The
corresponding {\it I--V\/} characteristics are displayed in Fig.~\ref{Fig4}(b).
Four current peaks are observed (labelled 1--4) which supply an emission
current 1.8 times greater than that predicted from conventional
Fowler--Nordheim tunnelling. In addition to these symmetric current peaks,
we find two highly asymmetric peaks, labelled FN$^1$ and FN$^2$ in
Fig.~\ref{Fig4}. The presence of the asymmetric peaks becomes
particularly apparent when the {\it I--V\/} characteristics of
Fig.~\ref{Fig4}(b) are displayed on a semi-logarithmic plot, as shown in
Fig.~\ref{Fig4}(c), where discontinuities aligned with peaks FN$^1$ and
FN$^2$ are clearly observed. It is possible to match
the experimentally observed field--emission resonance peaks to our theoretical 
results presented earlier.
Fits to the line shape of peaks FN$^1$ and FN$^2$, shown by the solid lines
in Fig.~\ref{Fig4}(c), yield pre-exponential factors of $V^{1.74}$ and
$V^{1.704}$, respectively. This is very close to the theoretically expected
power law $V^{5/3}$  predicted for resonant tunneling from the 3D
substrate states. Furthermore, the Lorentzian line shape of resonance peaks 
labelled 1--4 [see Fig.~\ref{Fig4}(a)] is in good agreement to our 
theoretically predicted lineshape corresponding to
resonant tunneling from the S--N accumulation layer. The excellent agreement 
between experiment and theory is a strong indication that resonant--tunneling
transport as illustrated in Fig.~\ref{Fig1} indeed occurs in our samples~\endnote
{The fact that our calculations were performed in the zero-temperature limit
while measurements were taken at room temperature does not affect this
conclusion. In particular, the predicted power law exhibited by the asymmetric
peaks will hold as long as the IV curve shows the cut-off behaviour characteristic
for resonant tunneling. However, tunneling from the 2D quantum-well states at the
N--S interface should be strongly affected because, in our samples, their confinement
energies are comparable to the room-temperature energy scale. This indeed explains
the broad shape and small height of the symmetric peaks. Their Lorentzian
form will be preserved, however, until such peaks are completely washed out.}.
Additional numerical calculations based on previously described
methods~\cite{lito:jvac:96,huang:jap:95, vat:jap:97,gonch:mat:03,wang:prb:05}
and further modelling of sample electrostatics could yield the bound--state
energies and their life--time broadening but these extended studies are beyond
the focus of this work.

In conclusion, we have studied field emission from nanostructured surfaces
that is strongly affected by resonant tunneling from both a 3D substrate
and from 2D accumulation layers. Universal features in the current--voltage
charateristics are derived theoretically and confirmed experimentally, and allow 
for identification of resonant--tunneling peaks with specific microscopic tunneling
mechanisms. While the experimental part of this study considers resonant
tunnelling from silicon nanostructure field--emission cathodes, the analytical
expressions derived are quite generic and are applicable to a range of
semiconductor structures used in the study of field--emission phenomena.

This work was performed under research contracts to the New Zealand
Foundation for Research, Science and Technology (C05X0008, `Advanced
industrial materials at the nanoscale').

\end{document}